# Temporal dynamics of diode-pumped circulation-free liquid dye lasers

A. Hamja[1,a)], S. Chénais[1], S. Forget[1]

[1]*Université Sorbonne Paris Nord, Laboratoire de Physique des Lasers (LPL), CNRS, UMR 7538, F-93430, Villetaneuse, France*

A highly stable diode-pumped circulation-free liquid dye laser in a vertical external cavity is reported. The design is simple (no fabrication process step required, no fluid circuitry), compact (~ cm sized), and cost-effective. An optical efficiency of 18% with a M² of 1 are reported, with an excellent photostability – no efficiency drop was seen after 1.4 million pulses at 50 Hz, a value comparable to flowing systems and much higher than values achievable with organic solid-state lasers. We show that thermal effects are central in the stability and also on the dynamics of this laser. The laser build-up and shutdown dynamics are studied in detail for different pump pulse durations/repetition rates; they reveal a pulse shortening with increasing pump pulse duration and repetition rate that are shown to be due to thermal lensing diffraction losses. This laser structure offers a very convenient and simple platform for testing or harvesting solution-processable gain materials.

## I. Introduction

More than 50 years following their first demonstration, liquid dye lasers [1] are still widely used in spectroscopy [2], medicine [3] or sensing [4], mainly in virtue of their intrinsic wavelength tunability. However, they are expensive and bulky, and often considered to be complicated to handle and maintain because of the dye solution circulation. Solid-state alternatives to liquid dye lasers have hence emerged, notably supercontinuum sources but they have limited spectral power and low temporal coherence. The advent of organic semiconductor lasers [5] has opened new horizons for compact/integrated lasers, and notably triggered intense research towards electrically-pumped organic laser diodes; [6] however it has not lead up to date to commercial products due to low photostability and modest output power or brightness [5,6] Liquid dye lasers, provided they could be made cheaper, smaller, and more user-friendly, would have a major interest in the field of tunable visible coherent sources.

This evolution of dye laser technology has today become possible thanks to the development of high-power blue/green laser diodes, and especially their spectacular recent price drop — indeed, multi-Watt 450 nm diodes can now be found at prices below

[a)] Electronic email: mdamir.hamja@univ-paris13.fr

100 US$. Consequently, "diode-pumped dye laser" reports have emerged in the recent literature.[7–9] It is worth noticing that albeit these reports have dealt with traditional laser dyes (rhodamines, coumarins, pyrromethenes…), the concept is also transferrable to novel solution-processable gain materials, *e.g.* soluble organic semiconductors or colloidal quantum dots.[10]

Diode-pumped liquid dye lasers have been demonstrated with both circulating[11,12] and non-circulating[7–9] dye solutions. In the first category, Stefanska *et al.*[11,12] adapted a traditional commercial dye laser to a diode pumping configuration and obtained Continuous-Wave (CW) single-mode lasing. However, one main limitation of liquid dye lasers is the requirement of a dye circulation system. Using dye solutions without circulation represents a huge gain in terms of practicability and compactness. But as molecules are not removed rapidly, the pulse duration and repetition rates are limited to some extent (as discussed in this paper). The pulse duration is typically limited to a few hundreds of ns [13] by triplet piling up ; triplet quenchers can substantially increase this value [14] but do not enable however reaching CW operation.

Recently, Coles *et al.* [15] have shown CW lasing in a circulation-free dye laser, based upon the idea that a tiny excitation spot volume (~10 μm³) enables diffusion of photogenerated triplet states / photobleached molecules fast enough outside of the mode volume. This interesting approach required a small pump spot size and hence did not use a diode as the pump source, as it is actually hard to obtain such tiny spots with high-power diode lasers. Moreover, in addition of a highly unstable laser output intensity, the efficiency was low (around $10^{-6}$) and fast photodegradation (within 3 minutes) was reported, which makes this concept not readily transferable for applications.

Circulation-free diode-pumped dye lasers were reported by Burdukova *et. al.* in pulsed mode, with 200-ns long pump pulses at low repetition rates (2 Hz). Lasing was demonstrated in transverse pumping [7], with a wide tunability [8] and high optical efficiency (~ 16.9% using DCM [9]). In these works, the cavity was long (more than 40 cm) and the output was not a pure $TEM_{00}$ fundamental cavity mode [8,9]. No degradation study was reported.

In this paper, we report on an efficient (18% optical efficiency), compact (cavity length < 1 cm), high-beam quality ($TEM_{00}$), diode-pumped circulation-free liquid dye laser, showing an unexpectedly high long-term stability (no drop observed after more than $10^6$ pulses at 50 Hz) that is comparable with flowing dye lasers and much better than what is achieved in solid-state media. We show that this unexpectedly high stability obtained without circulation cannot be attributed to diffusion alone but is presumably related to the existence of strong thermal gradients in the medium. We then study laser performance in a large range of repetition rates and pulse durations, a possibility that is offered readily by laser diodes, but was impossible with traditional gas or solid-state lasers that used to be employed for dye laser pumping in the past. The limitations of our system could then be investigated through a separate study of pulse duration and repetition rate influence on performance and stability.

The cavity design is very comparable to state-of-art external-cavity solid-state dye laser [16] but is easier to implement as the thin film fabrication step is skipped. We used a solution of DCM ([2-[2-[4-(dimethylamino) phenyl] ethenyl]-6-methyl-4H-pyran-4-ylidene] propanedinitrile) in ethanol as the laser gain medium, simply dispensed into a 1-mm-long (light path) cuvette without dye flow. We show that the deleterious role of triplet absorption and thermal lensing are the main limiting factors in terms of pulse width and repetition rate, respectively, but with limits that are pushed beyond what is currently accessible with solid-state media.

The influence of thermal lensing effects is here investigated by an estimation of Temperature-dependent refractive index change using laser interferometry. Finally, we present future perspectives of our work in the direction of entering longer pulse (towards CW/quasi-CW) and higher repetition rate operation.

## II. Experimental setup

Two 445 nm InGaN diodes (5W CW each) were used as the pump source, matching the high absorption of DCM dye at this wavelength. They were driven in pulsed mode using a PCO-7120 diode driver (Directed Energy Inc.). Diode beams were found to be elliptical and slightly astigmatic with $M^2$ ~ 13 and 1 along two orthogonal axes. Though astigmatism can be corrected using a cylindrical lens pair of proper focal lengths, this technique does not guarantee beam circularization while a circular pump is highly desired for longitudinal pumping to produce a diffraction-limited $TEM_{00}$ beam with high efficiency. Moreover, the setup becomes strongly sensitive to the alignment of the cylindrical lens pair. To avoid these issues, the two diode beams were first polarization-coupled, using a half-waveplate and a polarizing beam splitter (PBS) and then injected into a $\phi = 200$ μm core diameter optical fiber with a numerical aperture of NA = 0.22 (Figure 1). The goal was to achieve a circular pump at the fiber outlet and suppress the inherent astigmatism of diode beams, at the expense of a small power loss (15% at fiber injection) and a more significant brightness loss, which is not too problematic as far as the gain medium remains thin.

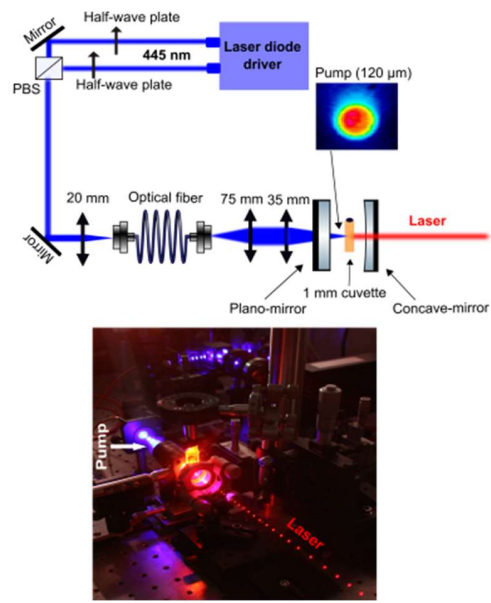

FIG. 1. Schematic diagram of experimental set-up (PBS: Polarization beam splitter). Inset: Long exposure time photo of the operation of diode pumped dye laser.

The beam was then focused by a pair of lenses onto a 120-μm-in-diameter spot inside a 1-mm-path length cuvette containing $1.98\times10^{-4}$ M solution of DCM in ethanol (76% absorption @ 445 nm). According to our previous findings, choosing higher concentration would cause the pump to be absorbed at the beginning of the cuvette hence letting the rest of the solution unpumped, which brings additional losses inside the cavity as DCM is not a perfectly true 4-level system. The cuvette facets were set at exact right angle of pump axis, inside a Plano-concave cavity consisting of a dielectric plane mirror (Transmission > 80 % @ 445 nm, Reflectivity R > 99.9% @ 600-660 nm) and a concave output coupler with Reflectivity chosen among ~ 99.9% / 98% / 94% / 90% @ 600-660 nm. Output coupler radius of curvature was chosen to be 100 mm based on mode overlap calculation between pump and cavity mode in a short cavity (< 1 cm). This choice also gave the possibility to comfortably extend cavity length up to 90 mm in order to insert intracavity elements for applications like frequency tuning without significantly disturbing the stability criteria and mode matching between cavity and pump mode. Two fast photodiodes were employed to simultaneously record temporal profile of both pump and laser using 1 GHz oscilloscope. Laser energy and stability was measured by a Gentec pyroelectric laser energy meter.

### III. Laser performance

Lasing was observed with a well-defined threshold around 23 kW/cm$^2$ when using high-reflectivity output coupler (Figure 2) in a 7-mm long plano-concave cavity. The optimal optical-to-optical conversion efficiency was found to be 18% (Figure 2 inset) with the 98% reflectivity output coupler, a value comparable with the value reported in [9].

In this design, the gain medium is pumped longitudinally (end-pumping), which has the major advantage to enable lasing in a single cavity spatial fundamental mode. This was confirmed by measuring the M$^2$ of the laser after a 150 mm lens outside the cavity (Figure 3). An M$^2$ value of 1.04 was extracted from fitting experimental data using the equation given as inset of Figure 3, indicating a pure TEM$_{00}$ diffraction-limited beam. In all the characterizations (Figure 2), we chose a pump pulse duration of 200 ns for comparison with comparable works in the literature, but we also obtained 17% and 14% optical efficiencies using longer pump pulses, 500 ns and 1 μs respectively (Figure S1 in supplementary information): this kind of pulse duration is rarely investigated but studying laser dynamics in this regime is a precious tool for probing the importance of triplet state piling up in laser performance.

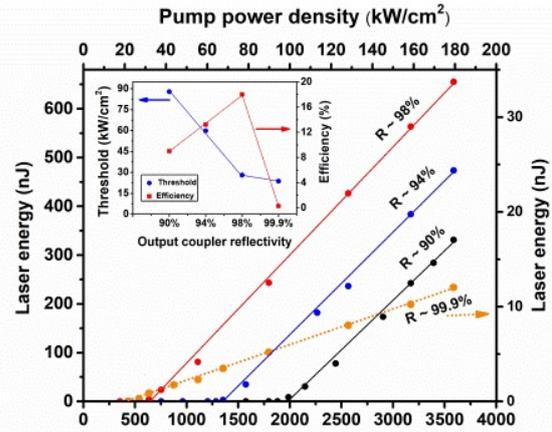

FIG. 2. Laser energy for different reflectivity (R) output couplers as a function of absorbed pump energy (bottom axis) & pump power density (top axis-measured from absorbed pump energy) for 200 ns pump.

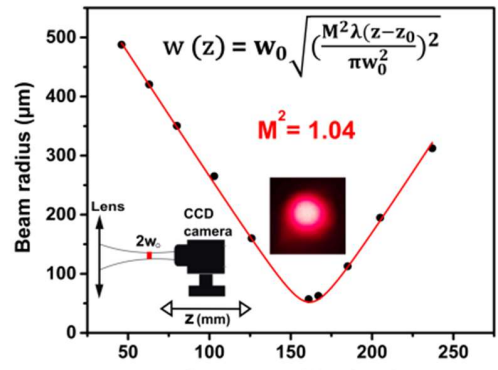

FIG. 3. Beam radius (μm) as a function of camera position (z) after f =150 mm lens (inset). Beam radius is minimum at waist (w$_0$). Experimental data (black dots) have been fitted (red line) using equation given as inset.

The evolution of laser pulse temporal profile with increasing pump power is shown in Figure 4 in a low-threshold cavity (highest output coupler reflectivity R > 99.9 %). It is important mentioning that all temporal profiles given in this paper (Figure 4, 5, 7, 8, S1, S4) are averages realized during several pulses (512 traces) and are identical from one pulse to another; in other words they do not reflect any permanent degradation of the medium but only a temporary lasing loss that is recovered at the next pulse (here the repetition rate is 10 Hz). We observe that under long pulse pumping (here 900 ns) there is almost no decrease in the laser intensity within the pump duration even under intense excitation of 153 kW/cm². It is also noticeable that the time delay between the onsets of pump and laser pulse, which is related to oscillation build-up time inside cavity, varies with pump power density. This build-up time depends on cavity length, pump power density and cavity loss[17]. For any laser having fixed cavity length and loss parameters (output coupler loss), build up time will be a function of pump power only. It is obvious in Figure 4 since the build-up time is higher (~150 ns) at lasing threshold and gradually decreases while increasing pump power.

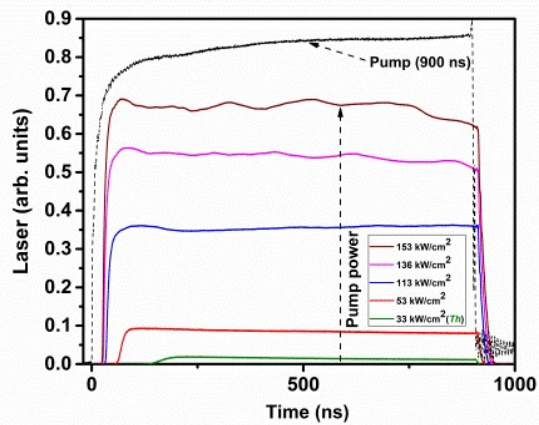

FIG. 4. Laser temporal profile using R~99.9% output coupler at different pump power densities under 900 ns/10 Hz pump. Build-up time decreases with increasing pump power

This stable ~1-μs pulse is quite different from what is typically observed in solid-state organic lasers, where lasing usually terminates after only a few tens of nanoseconds due to significant accumulation of long-lived excited triplet excitons through intersystem crossing (ISC). To illustrate this difference, we fabricated a 17-μm thick layer of DCM doped in PMMA (doping level 0.4% in weight, to ensure 83% of absorption, comparable to the cuvette absorption) directly spin-coated on the flat HR mirror closing the laser cavity. The absorption of liquid and solid-state media were thus very comparable, as were the pump parameters and cavity configurations. The pulse dynamics are shown in Figure 5. The pulse duration of liquid dye laser is here slightly different (not as "flat" as what is shown on Figure 4) because we used here R = 98% for the output coupler, corresponding to the optimal efficiency and output energy, instead of 99.9%. In this case, the additional amount of losses (2%) shortens the pulse temporal profile (see Figure S2 in the supplementary information for illustration) because as the laser threshold is higher, the singlet population is also higher and generates more triplet excitons through ISC. The full width at half maximum (FWHM) of the solid-state dye laser was found to be 240 ns whereas the liquid dye laser sustained oscillation up to more than 1μs. The reason why longer pulses are obtained in liquid state with no circulation is still under investigation, but can be due to concentration issues limiting the amount of Singlet-Triplet interaction in liquid experiments, or linked to self-diffusion of some triplet states outside of the gain medium, which might be facilitated by the pump-induced temperature gradient (see discussion at the end of the paper).

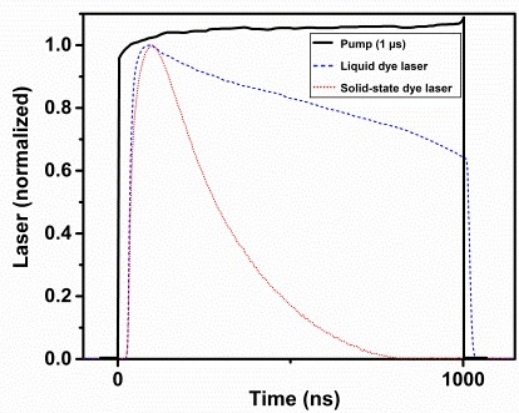

FIG. 5. A comparison between pulse duration of liquid and solid-state dye laser with same pump parameters and comparable absorption of solid-state and liquid gain medium

At last we investigated photodegradation of the device, that is the evolution of laser energy versus the number of pulses (Figure 6). It was also compared with the degradation characteristics of the solid-state dye laser (see above) under the same pumping conditions (1 µs pulse duration, 113 kW/cm²). The liquid dye laser showed no sign of degradation when operated at 50 Hz repetition rate up to around 8 hours (~ 1.4 million pulses). Meanwhile, the solid-state material was completely degraded after only 1000 pump pulses (the laser has then to be operated at 1 Hz to let enough time for measurement). This point illustrates the major advantage of using circulation-free liquid cuvettes instead of solid materials. To explain such a violent difference, one can first highlight that if the absorbed pump power is similar, it is not the case for the absorbed pump density which is much higher in the film, where hence a higher temperature rise is expected, which may speed up photobleaching. However, the absence of a downward trend suggests that the medium is fully replenished between two pulses, which might be due to self-diffusion of photobleached molecules outside of the pumped volume. Following this scenario, it is possible to estimate an order of magnitude of the time needed to escape the pumped volume based on simple diffusion equation[15]:

$$\tau_D = \frac{V^{2/3}}{6D}$$

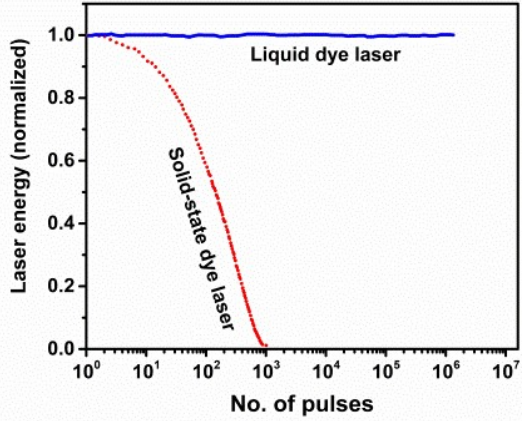

FIG. 6. Comparison of degradation of a liquid vs solid-state dye laser at 113 kW/cm² pump (1 µs) power density. Liquid dye laser was operated at 50 Hz up to approximately 8 hours. Solid-state dye laser was operated at 1 Hz until fully degraded

Where $\tau_D$ is a typical dye molecule diffusion time inside the pumped volume, V is the gain effective volume (= pump area × cuvette length) and D is the self-diffusion coefficient of a dye molecule in ethanol (D = 4.10⁻⁶ cm²/s [15]). This rough estimate leads here to $\tau \sim 21\ s$, which hardly explains why emission is so stable at 50 Hz. We conclude that thermal effects must play the major role in this surprisingly high emission stability. Because localized pumping induces a thermal gradient (see next part), a convective flux is created that carries photobleached molecules outwards, and hence reduces their dwelling time inside the gain volume.

### IV. Laser photo-physical study

In the previous section, we presented laser characteristics at low repetition rate (10 Hz or 50 Hz) and pulse durations up to 1 µs. However, in many applications, high repetition rate and or long pulses ("quasi-CW") laser is desirable. Therefore, we explored the laser behavior under longer pump pulse and higher repetition rate operation. To do so, repetition rate was first fixed at 10 Hz and pump pulse duration was increased up to 10 µs keeping pump peak power fixed. Unlike what was seen in Figure 4, laser pulse shortening was observed with

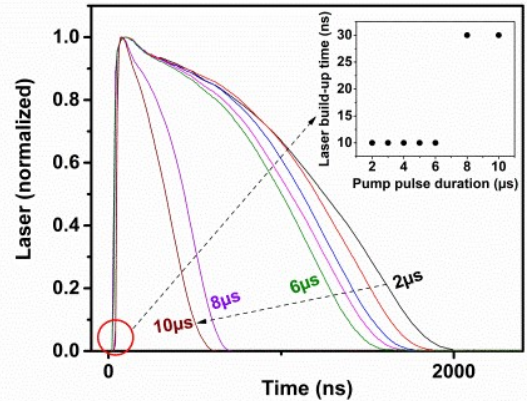

FIG. 7. Laser temporal profile as a function of pump pulse duration. Peak pump power (82 kW/cm²) and repetition rate (10Hz) were fixed. Inset: Laser build-up time (ns) vs pump pulse duration (µs).

increasing pump pulse duration (Figure 7). Going from 2 µs – 6 µs pump, the laser pulse width decreased almost linearly (FWHM from 1218 ns to 966 ns respectively). This is not due to permanent degradation of dye since the traces were obtained through averaging (over 512 traces) and were stable from pulse to pulse. However, beyond 6 µs, the shortening was much more significant, down to

234 ns. It is instructive to relate this to a change in the oscillation buildup time, as plotted in inset of Figure 7. A sudden buildup time increase (from 10 to 30 ns) is observed at 6 µs duration, which indicates, as pump peak power and cavity mirror losses are fixed, that the loss level has increased at this point. Causality dictates that these losses cannot be directly related to the duration of the pump but are due to some physical effect linked to previous pulses. This effect has a relaxation time longer than 100 ms (corresponding to the period of pulse excitation train). A related experiment (Figure 8) consists in increasing the pump repetition rate beyond 10 Hz, while fixing other pumping parameters (peak power density still fixed at 113 kW/cm², pulse duration fixed at 1 µs). Here, we observed an increase of buildup time for repetition rates above 30 Hz (indicating higher losses), and a consistent decrease of lasing duration. No more lasing was observed at and beyond 80 Hz (Figure S3). This once again proves the existence of a source of losses that is both pump-power dependent but preexisting when the pump pulse hits the material.

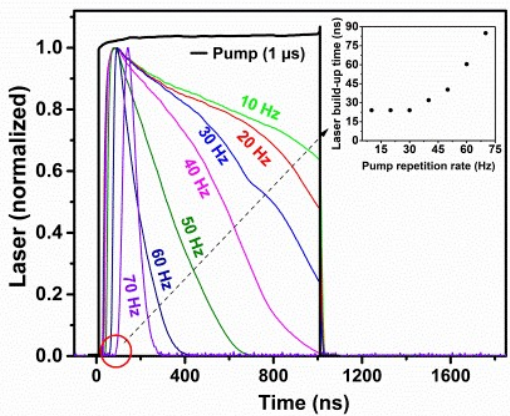

FIG. 8. Laser temporal profile as a function of pump repetition rate (Hz). Peak pump power was 113 kW/cm². Inset: Laser build-up time (ns) vs repetition rate (Hz).

This indicates that this loss is most probably of thermal nature. Indeed, the physical origin of short pulse lasing under CW pumping is well known in organic systems and is related to triplet states (triplet absorption and singlet-triplet annihilation. Typical pulse widths are in the sub-µs range in the absence of triplet quencher. However, it is here unlikely that the shortening observed when pulse duration or repetition rate is varied could be directly linked to an accumulation of triplet population between pulses, as the involved relaxation time (>100 ms for 10 Hz) is three orders of magnitude higher than the triplet lifetime (31±5 µs)[18] of DCM in solution. We also rule out the possibility of a permanent photoisomerisation or photobleaching scenario as the pulse durations are identical from pulse to pulse.

It is obvious that when the repetition rate is increased at constant pump energy per pulse (or when the pulse duration is increased at constant peak power and repetition rate), the average power is higher, and so is the thermal load inside the gain medium. After pump-induced heating during one pulse, the temperature relaxes down to room temperature with a typical timescale[19]:

$$\tau_T = \frac{w^2}{4D_T} = \frac{w^2 \rho c}{4K_c}$$

Where w is the pump radius (60 µm), $D_T$ the thermal diffusivity, ρ is the density, c is the specific heat and $K_c$ the thermal conductivity. From thermal values of ethanol ($K_c = 0.171 \text{W.m}^{-1}.\text{K}^{-1}$; $\rho = 789 \text{ kg.m}^{-3}$ ; $c = 2.46 \text{ kJ.kg}^{-1}.\text{K}^{-1}$ ), $\tau_T$ is ~10 ms. Generally, thermal effects (thermal lens dioptric power for instance) are proportional to the *average* absorbed pump power. However this is true in pulsed mode only if the temperature has enough time to relax to the initial temperature (room temperature) between two pulses, otherwise the temperature rise will build up pulse after pulse until reaching a novel average value that is dependent on the repetition rate[20]. In the experiment of Figure S4 (supplementary information), we maintained an average power of 0.57 mW by choosing different combinations of pump pulse duration and repetition rates; we observed that the pulse dynamics were actually very different, with much shorter pulses (indicating higher losses) at high repetition rates. This is a confirmation that the thermal timescale is here in the range of few tens of ms (corresponding to repetition rates between 10 and 100 Hz).

In the following, we investigate the physical origin of thermally-induced loss. We have seen previously that the very high stability could be explained by convective flux resulting from the pump-induced thermal gradient. This thermal gradient induced by localized heating also creates a negative thermal lens, as the temperature coefficient of refractive index dn/dT is large and negative in ethanol (-3.9×10⁻⁴/K)[21]. This thermal lens (potentially with aberrations) in turn creates diffraction losses[22,23] in the medium that add up to other losses (mirror losses, reabsorption losses and triplet absorption).

To demonstrate the existence of a significant thermal lens inside the medium, we probed the pumped area of the gain medium using a Michelson interferometer employing a red He-Ne laser (not absorbed by dye) with the liquid cuvette in one arm of the interferometer (Figure S5). At high repetition rates (80 Hz or 200 Hz), dark and bright fringes started to appear in the middle of probe beam indicating a distortion of the beam wave front due to the change in refractive index at the pump area location. To further quantify our findings, a power meter was placed at the center of the probe beam and diode power was increased for different repetition rates (10 Hz, 80 Hz & 200 Hz). In Figure 9 it appears that at 10 Hz, almost no thermal effect is observed, while at 80 Hz a phase variation of π (from a bright to a dark fringe) is observed for 100 kW/cm² pump fluence. At 200 Hz, a 2π dephasing is achieved for 113 kW/cm² of pump fluence only. Calculating the corresponding variation of the optical path gives a change in refractive index of $6.42\times10^{-4}$ which corresponds to a temperature gradient of 1.64 K.

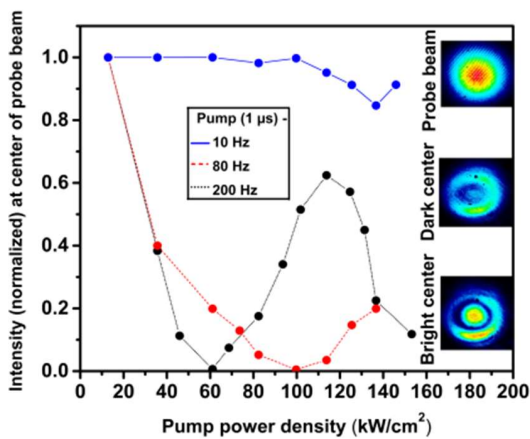

FIG. 9. Variation of the probe beam intensity (center of the fringe pattern) versus pump power density, for different repetition rates (10 Hz, 80 Hz and 200 Hz)

This study provides a strong indication that the major limitation of circulation-free dye lasers in terms of repetition rate is thermal lensing and the resulting induced diffraction loss. In order to increase the repetition rate above 80 Hz for practical applications, thermal load must be reduced, which can be done with diode lasers operating at lower pulse durations. In order to show the potential of this approach up to the kHz level, we used here 25 ns frequency doubled laser (532 nm, Harrier Quantronix) with 1.6 MW/cm² excitation density. With such pumping strategy, thermal load on gain medium was much lower due to short pump and low absorption of DCM at 532 nm (21% in our case using a 1 mm cuvette). In this case, lasing at 1 kHz was obtained and no degradation was recorded up to 18 million pulses (5 hours) and more (Figure S6). This highlights the interest of a compact liquid cuvette configuration for many applications operating at the kHz range.

**V. Conclusion**

Diode-pumped circulation-free liquid dye laser has been demonstrated with a pulse-to-pulse stability that is much higher than with current solid-state systems, while being comparable to current state-of-art organic thin film based solid-state lasers in terms of efficiency (18%) and beam quality (M²=1) without compromising simplicity, compactness and cost. The unexpectedly high photostability is supposed to be due to a convective diffusive transport of photobleached molecules outside of the pumped region, ensuring a complete replenishing of the gain medium between each pulse. Thermal effects were also proven to be central in order to explain the behavior of the laser under long pulse operation or high repetition rate. Indeed, 1 µs stable and efficient lasing has been achieved at 10 Hz but could not be increased for longer pump pulses (up to 10 µs) or higher repetition rate (up to 80 Hz); while the pulse duration is fundamentally limited by triplet piling up, we observed a pulse shortening at longer pulse durations and higher repetitions rates which can only be explained by a very-long characteristic time phenomenon, which we attribute to thermal lensing diffraction losses. Using shorter pulses for pumping, it was possible to reduce the thermal load and obtain lasing up to the kHz level. The device demonstrated here is very simple and can be further improved. Fundamental limit for pulse duration can be overcome with triplet quenchers, while thermal issue can be mitigated, even at high repetition rates, by decreasing the thermal load (e.g. shorter pump pulses), modifying the cavity, or replacing ethanol with another solvent having better thermal conductivity. Such a simple and stable laser can be an alternative to bigger systems for applications where a tunable pulsed source is required in the visible spectrum; but it can also be employed for a quick and easy characterization of novel gain media in the emerging field of solution-processable excitonic gain materials.